\documentstyle[12pt]{article}   
\newcommand{\Pt}{P_{\tau}}
\begin{document}
\title{Electrostatic Bender Fields, Optics, Aberrations, with Application to the Proton EDM Ring}

\author{R. Baartman, TRIUMF}
\date{Dec.\ 2013}

\maketitle
\begin{abstract} 
Electrostatic bender optics are derived up to second order (third order in fields and the Hamiltonian) and applied to the proposed EDM proton ring. The results for linear optics agree with those already presented by V.\ Lebedev (Nov.\ 18, 2013). Second order optics is not sensitive to the shape of the fringe fields and formulas are given. It is shown that the proposed electrode shape that linearizes the vertical electric field is no advantage to this order.
\end{abstract}

\section{Introduction}
The general electrostatic bender has a given curvature in the median plane (arbitrarily assumed here to be horizontal), and, to control the motion in the vertical direction, also has curvature in that direction. It is assumed that the two curvatures are constant radius, but different, so the equipotentials are sections of a torus. Lebedev\cite{leb} has mentioned that the electrodes can be shaped in such a way that the vertical field is linear, i.e., and so has no $xy^2$ coupling term. This shape will also be considered. 

Though the electrodes would be continuous in the vertical plane, they are necessarily truncated in the horizontal plane to allow the particles to enter and exit the field. Thus the radius of curvature along the reference trajectory changes continuously from infinity outside the field region to a finite value inside. The electric potential of such a truncated torus is derived to third order in deviations from the reference trajectory; the third order term contains the second derivative of the curvature function thus enabling analysis of lowest order fringe field effects. The relativistically correct Hamiltonian is derived also to third order. A canonical transformation is found which simplifies the Hamiltonian by eliminating derivatives of the curvature function. This allows an analysis of the linear optics and simple formulas for second order aberrations arising both from the curved geometry and the fringe fields. The linear optics agree with known; the second order aberration formulas were presented previously for the non-relativistic limit\cite{snowmass}.

\section{The Potential}
In a general curvilinear system the Laplacian of a potential $V(q^1,q^2,q^3)$ is given by
\begin{equation}\label{curv1}
\nabla^2 V=  \frac{1}{h_1h_2h_3}\sum_{i=1}^3\frac{\partial }{\partial q^i}\left(\frac{h_1h_2h_3}{h_i^2}\frac{\partial V}{\partial q^i}\right),
\end{equation}
where the $h_i$ are the Lam\'{e} coefficients and in general are functions of the $q^i$.. 

\subsection{Toroidal coordinates}
We start with a curvilinear coordinate system that conforms to the shape of the equipotentials: the coordinate along the intended reference orbit we call $q^3=s$, the radially outward coordinate is $q^1=x$, and the vertical is $q^2=\tilde{y}$. The horizontal curvature (reciprocal of radius) is $h(s)$, and the vertical curvature is $k$, a constant. In such a system, the Lam\'{e} coefficients are $h_1=1,\ h_2=1+kx,\ h_3=1+h(s)x$.

But in this system, the potential has no dependence on $\tilde{y}$, the direction orthogonal to $x$ and $s$. We can therefore expand to third order as\footnote{This potential is dimensionless. Multiply by $\beta c(B\rho)=\beta cp_0/q=\beta^2\gamma mc^2/q$ to get the potential.}
\begin{equation}
\tilde{V}(x,s)=a_{10}(s)x+a_{20}(s)\frac{x^2}{2}+a_{30}(s)\frac{x^3}{6}
\end{equation}
We know the electric field on the reference orbit imposes the curvature $h$, so $a_{10}(s)=h(s)$. We can find the other coefficients from Laplace's equation and eqn.\,\ref{curv1}:
\begin{equation}\label{pot1}
\tilde{V}(x,s)=hx-\frac{1}{2}h(k+h)x^2+\frac{1}{6}\left[2 h \left(k^2+k h+h^2\right)-h''\right]x^3
\end{equation}

\subsection{The Potential in Frenet-Serret Coordinates}
The Frenet-Serret coordinate system is the usual one of accelerator physics. It has no curvature in the $y$-direction, so the factor $h_2=1$ in eqn.\,\ref{curv1}. The potential in the median plane is the same in both systems, we need only add the following two more terms allowed by the symmetry: $a_{02}(s)y^2/2$, $a_{12}(s)xy^2/2$.
\begin{equation}
 V(x,y,s)=\tilde{V}(x,s)+\frac{1}{2}a_{02}(s)y^2+\frac{1}{2}a_{12}(s)xy^2.
\end{equation}

We again use the curvilinear Laplacian (\ref{curv1}), but this time with $h_2=1$, set it to zero and solve term-wise in the expansion to find $a_{02}$ and $a_{12}$:
\begin{equation}\label{pot2}
 V(x,y,s)=hx-\frac{h}{2}(k+h)x^2+\frac{hk}{2}y^2-\frac{kh}{2}(2k+h)xy^2+\frac{1}{6}[2h(k^2+hk+h^2)-h'']x^3
\end{equation}

Lebedev's\cite{leb} linear vertical electric field (VL) potential also can be augmented to include lowest order fringe field effects and still satisfy Laplace to third order. The result is slightly different from the one above:
\begin{equation}\label{YSpot}
 V(x,y,s)=hx-\frac{h}{2}(k+h)x^2+\frac{hk}{2}y^2+\frac{1}{6}[h^2(k+2h)-h'']x^3
\end{equation}

The reason for the difference is that potential \ref{pot2} is derived for electrodes whose vertical sections are exact circular arcs while the potential \ref{YSpot} was derived with the specific intention to zero the $xy^2$ term. We shall see how the dynamics of the two compare.

\section{Dynamics}
\subsection{General Hamiltonian for Electrostatics}
First, we must decide on the dynamical coordinates, especially the longitudinal (third) generalized momentum, which has been the source of considerable confusion. A common approach, used especially by those accustomed to only magnetic elements, is to let the third momentum be $\Delta p/p$. This is not the simplest approach, since when electric fields are included, it is not conserved. This means that when a particle enters the electrostatic element off-axis, it must receive a ``kick'' to get into the potential
field. This kicks $p$, but leaves $E$ unchanged. For an electrostatic bend of radius $A$, this kick is $\Delta p/p=\pm x/A$; the
upper sign is for entry, the lower for exit.

A better approach is therefore to let the third momentum be $E$. Since the fields are static, $E$ is conserved; no kicks are required.

For independent variable $s$, on a reference trajectory curving in the
$xs$-plane with curvature $h(s)=1/R(s)$, the Hamiltonian $H=-p_s$ is well-known, and I
will not derive it here:
\begin{equation}H=-(1+hx)\sqrt{\left({E-q\Phi\over c}\right)^2-m^2c^2-p_x^2-p_y^2}\end{equation}
We write $E=E_0+\Delta E$, and note that the large quantity under the
square root sign is 
\begin{equation}p_0^2=E_0^2/c^2-m^2c^2=(\gamma^2-1)m^2c^2=(\beta\gamma mc)^2,\end{equation} the
square of the reference momentum.

\begin{equation}H=-(1+hx)p_0\sqrt{1+{2E_0(\Delta E-q\Phi)\over p_0^2c^2}+\left({\Delta
      E-q\Phi\over p_0c}\right)^2-{p_x^2\over p_0^2}-{p_y^2\over p_0^2}}\end{equation}

Let us transform so that the third coordinate is not time, but a relative distance deviation w.r.t.\ the reference particle: i.e.\ from $(t,-\Delta E)$ to $(\tau,p_\tau)$ where $\tau\equiv s-\beta ct$, $p_\tau=\Delta E/(\beta c)$. The generating function is 
\begin{equation}
F(t,p_\tau)=(s-\beta ct)p_\tau
\end{equation} 
then the new Hamiltonian is $K=H+\partial F/\partial s=H+p_\tau$
\begin{equation}
K=p_\tau-(1+hx)p_0\sqrt{1+2\left({p_\tau\over p_0}-{q\Phi\over\beta cp_0}\right)+\beta^2\left({p_\tau\over p_0}-{q\Phi\over\beta cp_0}\right)^2-{p_x^2\over p_0^2}-{p_y^2\over p_0^2}}
\end{equation}
This cleans up considerably if we scale all momenta and the Hamiltonian by $p_0$: $P_x=p_x/p_0$, $P_y=p_y/p_0$, $P_\tau=p_\tau/p_0$, $\tilde{K}=K/p_0$, and introduce the scaled potential $V={q\Phi\over\beta cp_0}$:
\begin{equation}
\tilde{K}=P_\tau-(1+hx)\sqrt{1+2\left(P_\tau-V\right)+\beta^2\left(P_\tau-V\right)^2-P_x^2-P_y^2}
\end{equation}
This also makes the momenta accord with the more usual definitions, since, $P_x=x'$, $P_y=y'$, and outside the electric field, $P_\tau=\Delta p/p$.

This is the Hamiltonian of the electrostatic bend. It is exact.

\subsection{Linear Optics}
Linear optics arise from second order Hamiltonian terms. Expanding to second order we get for both the potentials  eqns.\,\ref{pot2}, \ref{YSpot}:
\begin{equation}\label{pot}
V=hx-h(h+k){x^2\over 2}+hk{y^2\over 2},
\end{equation} 
where $h=1/R_0$ and $k=1/R_y$ is the curvature in the non-bend direction. The first term, needed to ensure that the reference trajectory is $x=0$, yields the required electric field on the reference orbit:
\begin{equation}
{\cal E}= -{\partial\Phi\over\partial x}=-{\beta cp_0\over q}\,\left.{\partial V\over\partial x}\right|_{x=0}={\beta^2\over R_0}{E_0\over q}.
\end{equation}
In the non-relativistic limit, the electric field is twice the beam kinetic energy divided by charge and bend radius: $q{\cal E}=mv^2/R_0$.

The first order terms in the resulting Hamiltonian all cancel as they should or the reference orbit would not be an orbit, so when expanded to second order it is
\begin{equation}
\tilde{K}_1={P_x^2\over 2}+{P_y^2\over 2}+{\Pt^2\over 2\gamma^2}-{2-\beta^2\over R_0}\,x\Pt+{\xi^2\over 2R_0^2}\,x^2+{\eta^2\over 2R_0^2}\,y^2
\end{equation}
The parameters $\xi$ and $\eta$ are introduced as they parametrize the $x$ and $y$ focusing strengths: $\xi^2+\eta^2=2-\beta^2=1+\gamma^{-2}$, $\eta^2=k/h=R_0/R_y$, $R_y$ being the curvature radius in the non-bend direction. ($\eta^2$ is given the symbol $m$ in proton EDM group convention.) In the non-relativistic limit, for a cylindrical bend, $\xi=\sqrt{2},\eta=0$; for a spherical bend, $\xi=\eta=1$.

The transfer matrix is easily derived from this Hamiltonian $\tilde{K}_1$:
\renewcommand{\arraystretch}{1.5}
\begin{equation}
\left( \begin{array}{cccccc}   
\cos \xi\theta & {R_0\over \xi}\sin \xi\theta & 0 & 0 & 0 & {2-\beta^2\over \xi^2}R_0(1-\cos \xi\theta) \\
-{\xi\over R_0}\sin \xi\theta & \cos \xi\theta & 0 & 0 & 0 & {2-\beta^2\over \xi}\sin \xi\theta \\
          0 & 0 & \cos \eta\theta & {R_0\over \eta}\sin \eta\theta & 0 & 0 \\
0 & 0 & -{\eta\over R_0}\sin \eta\theta & \cos \eta\theta & 0 & 0 \\
-{2-\beta^2\over \xi}\sin \xi\theta & -{2-\beta^2\over \xi^2}R_0(1-\cos \xi\theta) & 0 & 0 & 1 &
R_0\theta\left[{1\over\gamma^2}-{\left(2-\beta^2\over \xi\right)^2}\left(1-{\sin \xi\theta\over \xi\theta}\right)\right] \\
          0 & 0 & 0 & 0 & 0 & 1 
\end{array}\right)\nonumber
\end{equation}
This matrix agrees precisely with that given by Lebedev\cite{leb}, except that the 56 element explicitly includes the velocity dependent part $1/\gamma^2$, allowing to read directly the phase slip factor for a solid electrostatic ring: $\frac{1}{\gamma^2}-\left(\frac{2-\beta^2}{\xi}\right)^2$. Moreover, for such a ring, the horizontal tune is $Q_x=\xi=\sqrt{2-\beta^2-m}$, vertical tune $Q_y=\eta=\sqrt{m}$, dispersion $=\frac{2-\beta^2}{2-\beta^2-m}R_0$.

I have used this matrix to find the first order optics of a ring with 14 benders with $\eta^2=m=0.199$ and separated by $0.834$ metre drifts: tunes, $\beta$-functions, dispersion and phase slip factor agree precisely with the results presented by Valeri Lebedev\cite{leb}.

\subsection{Second Order Aberrations}
Expanding to third order, the Hamiltonian for the toroidal electrodes becomes
\begin{eqnarray}
 \tilde{K}_2=\tilde{K}_1-\frac{\Pt}{2}\left(P_x^2+P_y^2+\frac{\Pt^2}{\gamma^2}\right)
-\Pt x^2 \left(\frac{kh}{2\gamma^2}+2\frac{h^2}{\gamma^2}\right)-\Pt y^2 \frac{kh}{2\gamma^2}+x\left(P_x^2+P_y^2+2\frac{\Pt^2}{\gamma^2}\right)h\nonumber\\
+xy^2 \left(-k^2 h+\frac{kh^2}{2\gamma^2}\right)+x^3\left(\frac{k^2h}{3}-\frac{kh^2}{6}\left(1+\frac{3}{\gamma^2}\right)-\frac{h^3}{6}\left(1-\frac{3}{\gamma^2}\right) -\frac{h''}{6}\right)
\end{eqnarray}

For the potential \ref{YSpot}, it is only slightly different:
\begin{eqnarray}
 \tilde{K}_2=\tilde{K}_1-\frac{\Pt}{2}\left(P_x^2+P_y^2+\frac{\Pt^2}{\gamma^2}\right)
-\Pt x^2 \left(\frac{kh}{2\gamma^2}+2\frac{h^2}{\gamma^2}\right)-\Pt y^2 \frac{kh}{2\gamma^2}+x\left(P_x^2+P_y^2+2\frac{\Pt^2}{\gamma^2}\right)h\nonumber\\
+xy^2 \left(\frac{(1+\gamma^2)kh^2}{2\gamma^2}\right)+x^3\left(\frac{k^2h}{3}-\frac{kh^2}{6}\left(2+\frac{3}{\gamma^2}\right)-\frac{h^3}{6}\left(1-\frac{3}{\gamma^2}\right) -\frac{h''}{6}\right)
\end{eqnarray}
Thus even though the latter potential was designed to eliminate the $xy^2$ term, it re-appears in the Hamiltonian. The reason can be traced to the factor $1+hx$ in front of the square root; coupled with the focusing term $\propto khy^2$ this results in a term $\propto kh^2xy^2$ which is in fact larger than the term that was eliminated. Thus it is not clear that special non-circular-section electrode shapes are warranted.

\subsection{Fringe Field Shape Effect}
The presence of the second derivative of the curvature in the Hamiltonian seems to suggest that the effect in this order of the fringe field can be minimized by shaping it. For example, if $h(s)$ is linear, its contribution is zero. However, this technique proves to be futile: there necessarily will be regions of large $h''$ at either end of such a linear fringe field, and these will have the same integrated effect as a fringe field with no linear part. To be sure, third (often called octupole) and higher orders do depend on the fringe field shape; in general, the smoother and more extended the fringe field, the better.

In fact, the second (often called sextupole) order effects of the fringe field are insensitive to the fringe field shape. This follows from the fact that the second derivative can be transformed away with a canonical transformation to a new horizontal transverse coordinate.

We transform $(x,P_x)\rightarrow (X,P_X)$ by means of the following generating function:
\begin{eqnarray}
 G(x,P_X,s)&=&xP_X+h'\frac{x^3}{6}-h\frac{x^2P_X}{2}\\
 X=x-\frac{1}{2}x^2h&\mbox{and}&P_x=P_X-xP_Xh+\frac{1}{2}x^2h'
\end{eqnarray}
The new  Hamiltonian is $\tilde{\tilde{K}}(X,P_X,y,P_y,z,P_z;s)=\tilde{K}+\frac{\partial G}{\partial s}$. 
\begin{eqnarray}
\tilde{\tilde{K}}=\frac{P_X^2}{2}+\frac{P_y^2}{2}+\frac{\Pt^2}{2 \gamma ^2}+X^2\left(-\frac{kh}{2}+\frac{h^2}{2}+\frac{h^2}{2 \gamma ^2}\right)+y^2\frac{kh}{2}-\Pt X\left(h+\frac{h}{\gamma^2}\right)+\label{lin}\\
\Pt^2X\frac{2h}{\gamma ^2}+\Pt \left(-\frac{P_X^2}{2}-\frac{P_y^2}{2}-\frac{k y^2 h}{2 \gamma ^2}+X^2\left(\frac{k h}{2 \gamma ^2}-\frac{h^2}{2}-\frac{5 h^2}{2 \gamma ^2}\right)\right)-\frac{\Pt^3}{2 \gamma ^2}+P_y^2 X h+\label{pin}\\
X y^2 \left(-k^2 h+\frac{k h^2}{2 \gamma^2}\right)+X^3 \left(\frac{1}{3} k^2 h-\frac{2}{3} k h^2-\frac{k h^2}{2 \gamma ^2}+\frac{h^3}{3}+\frac{h^3}{\gamma ^2}\right)\label{xin}\end{eqnarray} 
As $(x,P_x)$ and $(X,P_X)$ differ from each other only in higher order, the linear optics is unchanged. The Hamiltonian can be written as a sum of linear part $K_1$ (\ref{lin}), momentum-dependent nonlinear part $K_P$ (\ref{pin}), and spatial nonlinear part $K_Q$ (\ref{xin}). 

\section{Results for $K_Q$}
With no derivatives of $h$ appearing, we can now to a good approximation assume $h=1/R_0$, a constant, and $k=m/R_0$. Further, scale lengths to be in units of $R_0$. Then
\begin{equation}
K_Q=X^3 \left(\frac{(m-1)^2}{3}+\frac{1}{\gamma ^2}-\frac{m}{2 \gamma ^2}\right)+X y^2 \left(-m^2+\frac{m}{2 \gamma^2}\right)=0.792X^3+0.024Xy^2
\end{equation}
The numerical values are for the proton EDM ring ($\gamma=1.2481$, $m=0.199$).

The same transformation applies to the (VL) potential that is designed to eliminate the $xy^2$ term. In that case only $K_Q$ is different. The result is:
\begin{equation}
 K_Q=X^3 \left(\frac{1}{3}-\frac{5 m}{6}+\frac{1}{\gamma ^2}-\frac{m}{2 \gamma ^2}\right)+X y^2 \left(\frac{m}{2}+\frac{m}{2 \gamma ^2}\right)=0.746X^3+0.163Xy^2
\end{equation}

As stated, the non-toroidal electrodes that make the vertical field linear appear to be of no advantage.

\end{document}